\documentclass[11pt,onecolumn,journal]{IEEEtran}
\usepackage{graphicx,epstopdf}
\usepackage{stfloats}
\usepackage{csquotes}

\usepackage{subfigure}
\usepackage{cite}
\usepackage{amsmath}
\usepackage{tabularx}
\usepackage{footnote}
\usepackage{adjustbox}
\usepackage{amsthm}

\usepackage{tikz}
\usepackage[left=1in,right=1in,top=1in,bottom=1in,%
            footskip=.25in]{geometry}
\usetikzlibrary{%
    decorations.pathreplacing,%
    decorations.pathmorphing%
}
\usetikzlibrary{shapes.geometric, arrows}

\tikzstyle{process} = [rectangle, minimum width=3cm, minimum height=0.5cm, text centered, text width=7cm, draw=black,fill=gray!10]
\tikzstyle{process2} = [rectangle, minimum width=3.8cm, minimum height=0.5cm, text centered, text width=3.8cm, draw=black, fill=gray!10]
\tikzstyle{process3} = [rectangle, minimum width=3cm, minimum height=0.5cm, text centered, text width=3cm, draw=black, fill=gray!10]
\tikzstyle{process1} = [rectangle, minimum width=2.5cm, minimum height=0.5cm, text centered, text width=2.5cm, draw=black, fill=gray!10]
\tikzstyle{decision} = [diamond, minimum width=3cm, minimum height=0.5cm, text centered, draw=black, fill=green!30]
\tikzstyle{arrow} = [thick,->,>=stealth]
\begin{document}
\linespread{1.25}
\title{Near-Field Radiation Exposure Control in Slot-Loaded Microstrip Antenna: A Characteristic Mode Approach}
\author{Sandip~Ghosal,~ Arijit~De,~Raed M.~Shubair,~and Ajay~Chakrabarty
\thanks{S.~Ghosal, A.~De and A.~Chakrabarty are with the Department
of Electronics and Electrical Communication Engineering, Indian Institute of Technology, Kharagpur,
West Bengal 721302, India (e-mail: sgmw@iitkgp.ac.in).}
\thanks{Raed M. Shubair is with Research Laboratory of Electronics, Massachusetts Institute of Technology (MIT), Cambridge, MA 02139 USA (e-mail: rshubair@mit.edu).}}
\maketitle
\vspace*{-1em}
\begin{abstract}
Microstip antenna topology is commonly loaded with a narrow slot to manipulate the resonance frequency or impedance bandwidth. However, the tuning of the resonance frequency or impedance bandwidth results in the variation of the current and field distributions. In this regard, this work adopts the concept of characteristic modes to gain an initial understanding of the perturbation mechanism of the rectangular patch when loaded with a slot. The performance of microstrip antennas with finite ground plane is then studied using full-wave simulation. It has been found that the distribution of the induced current density is highly dependent on the orientation of the slot The incorporation of a narrow slot suppresses the nearby orthogonal eigen mode and, as a consequence, the radiation behaviour is  affected. Specifically, in the presence of biological tissues in the near-field region, both antenna input impedance properties and the realized gain are dependent on the slot orientation. Different examples are included for understanding the impact of slot loading on the energy absorption by biological tissues, by calculating the the specific absorption rate (SAR). The proposed analysis facilitates the design of  miniaturized antenna geometries for biomedical applications via systematic loading of narrow slots.
\end{abstract}
\newpage
\section{Introduction}
The last two decades have witnessed an exponential growth and tremendous developments in antenna technologies and techniques, and their associated applications, such as those reported in \cite{elayan_graphene-based_2018,elayan_graphene-based_2016,shubair_simple_1992,ibrahim_reconfigurable_2016,khan_pattern_2016,shubair_novel_2015,alharbi_flexible_2018,khan_novel_2018,khan_properties_2016,ghosal_characteristic_2018,khan_second_2017,khan_pattern_2017,shubair_complex_1993,kipp_numerically_1994,khan_triband_2018,karli_minituarized_2016,shubair_optimized_1998,shubair_technique_1992,shubair_full_1991,shubair_analysis_1992,shubair_combined_1993,shubair_complex_1993-1,karli_miniature_2016,shubair1993simple,shubair1993efficient,khan2016compact,shubair1993closed,al2005direction,nwalozie2013simple,che2008propagation,shubair2005robust,omar2016uwb,bakhar2009eigen,khan2017ultra,shubair2015novel,elsalamouny2015novel,ibrahim2017compact,shubair2005convergence,an2009compact,ray2008doa,hakam2016novel,hussein2016compact,al-ardi_investigation_2003,al-ardi_performance_2003} . Being an integral part of the cellular phone, antenna's near-field exposure on biological tissues is a major concern for the antenna designers. RF exposure induced by the GSM mobile phone has been studied using statistical modelling in \cite{wiart2000analysis}. A numerical technique has also been proposed to determine the RF exposure effects on a human head due to a cellphone equipped with a dual-band monopole-helix antenna \cite{bernardi2001power}. In a similar way, various statistical and experimental studies were carried out to assess the impact of mobile radiation on nearby biological tissues \cite{gati2009exposure} \cite{kuhn2012field} \cite{xu2018radio}. Recently, modal analysis technique has been adopted for predicting the frequency of maximum RF exposure \cite{lopez2016eigenmode} \cite{sandip_temc1}. Therefore, the antenna geometry should be designed in a way such that it maintains a certain threshold of RF energy absorption by the biological tissues commonly characterized by the SAR value \cite{zhu2016miniaturized} \cite{delong2018radiating}.\\

Antenna design in the presence of nearby biological tissues involves three main aspects. The first is to select the optimum frequency for maximum power transfer, as rigorously studied in \cite{poon2010optimal}. The second is to focus the radiated field or SAR distribution towards a desired direction \cite{poon2012electromagnetic}  \cite{bellizzi2017multi}. The third and last aspect, is to miniaturize the design while maintaining the desired radiation characteristics in the desired frequency of operation. Miniaturization of the design is achieved by incorporating electromagnetic band gap (EBG) structures \cite{fan2018miniaturized} or slot elements \cite{ghosal2013analysis} in the topology of the antenna. The latter technique is the focus of this report.\\

Slot loading of is a popular technique used by the antenna designers to manipulate resonance frequency \cite{khan2015microstrip}, enhance impedance bandwidth \cite{sze2000slotted}, or reduce side-lobe level
\cite{juyal2016sidelobe}. The principle behind slot loading technique is to manipulate the radiation behaviour which is directly controlled by the current density vector induced on the surface or volume of the radiating antenna. In this regard, a proper analysis is required to study the effect of slot loading on the radiated field and SAR. In the literature, slot loading phenomena has been discussed following three approaches: cavity modelling \cite{juyal2016sidelobe}, method of moment (MoM) technique \cite{pozar1986reciprocity}, and transmission line theory \cite{ghosal2013analysis}. Recently, a modal approach based on characterisitc mode analysis (CMA) has been employed to study the effect of slot loading \cite{SG2018APS} \cite{SG2018ANTEMPatch}. Characteristic mode theory was initially developed to determine the induced current vector using generalized eigen-decomposition of the MoM impedance matrix. If Galerkin's method is followed in the electric field integral equation (EFIE) formulation, the MoM impedance matrix becomes symmetric \cite{harrington1971theory}. Generalized characteristic equation of the MoM matrix provides the eigen-values and the characteristic current eigen-modes. As shown in \cite{harrington1971theory} and \cite{sandip_temc1}, these eigen-values  can predict the resonance frequency of the antenna structure. The amount of outward radiated power is also directly dependent on the eigen-modes and eigen-values. Unlike the conventional MoM technique, computation of these eigen parameters is independent of the excitation vector. It depends only on the operating frequency, and on the physical and geometrical properties of the radiating structure. Hence, with the knowledge of excitation-independent eigen-current distribution, one can perturb the radiating structure topology to achieve a desired radiation pattern. Such advantageous scopes of the CMA are found to be recently exploited in controlling the higher order modes \cite{lin2018method} and radar cross section \cite{zhao2018band} respectively. This report aims to extend the previously reported modal approach of \cite{SG2018APS} \cite{SG2018ANTEMPatch}, for perturbing the horizontal radiation in the near field of the radiating structure, such that the RF exposure on the on the nearby biological tissues is adjusted.\\

The report is organized as follows: Section II starts by describing the  characteristic mode analysis, then considers a rectangular geometry to study the perturbation of modal current density due to the loading of narrow slots. Section III explores the method further with full-wave examples and corresponding analysis of the SAR distribution Finally, conclusions are given in Section IV..

\newpage
\section{Modal Analysis} 
For the sake of initial understanding, let us consider a perfect electrically conducting (PEC) radiating element with no internal resonance. The outward unit vector normal to the surface of the antenna is denoted by $\hat{n}$. $\text{E}^i$ is the excitation field vector of the antenna. $\text{J}$ is the surface current density vector induced on the surface $S$ of the antenna. Following the field boundary conditions on the surface $S$ ~\cite{harrington1971theory},
 \begin{subequations}
\begin{align}
\hat{n} \times \text{E}^i&=-\hat{n} \times [j\omega A(\text{J}(r')+\nabla \phi(\text{J}(r')]\label{1a}\\
A(\text{J}(r')&=\mu \oint_{S}\text{J}(r')G(r,r') \,ds \label{1b}\\
\phi (\text{J}(r')&=\frac{-1}{j\omega\epsilon}\oint_{S} \nabla'\text{J}(r')G(r,r') \,ds \label{1c}\\
G(r,r')&=\frac{e^{-jk|r-r'|}}{4\pi|r-r'|}\label{1d}
\end{align}
\end{subequations}

The permittivity, permeability and wave-number of the free-space medium are represented by $\mu$, $\epsilon$ and $\emph{k}$ respectively. The integral equation of \eqref{1a}-\eqref{1d} can be converted into equivalent matrix form using the method of moment (MoM) technique~\cite{harrington1971theory} as,
\begin{equation}
[Z]_{M \times M}[I]_{M \times 1}=[V]_{M \times 1}\label{3}
\end{equation}

The MoM impedance matrix is denoted by $[Z]$ and the column vector $[V]$ consists of the weighting coefficients generated due to the inner product of the excitation vector $\text{E}^i$ and $M$ number of basis functions $\text{f}_m$.
Using the CMA of \cite{harrington1971theory}, the weighting vector $[I]$ of the induced current can be written as
 \begin{align}
[I]&=\sum_{m=1}^{M} \alpha_m [I_m],~\text{where}~\alpha_m=\frac{[I_m]^H[V]}{(1+j\lambda_m)[I_m]^H[R][I_m]} \label{5}
\end{align}
and the $m^{th}$ characteristic current mode $\text{J}_m$ will be
 \begin{align}
\text{J}_m&=\sum_{i=1}^{M} I_m(i)\text{f}_i(r'_m) , r'_m\in S \label{6}
\end{align}

The $m^{th}$ eigenvector $[I_m]$ and eigenvalue $\lambda_m$ can be obtained through the generalised eigen decomposition of the impedance matrix $[Z]$ as,
 \begin{align}
[X][I_m]=\lambda_m[R][I_m],~\text{where}~[Z]=[R]+j[X]
 \label{7}
\end{align}

If Galerkin's matching is followed in computing the $[Z]$ matrix, both $[R]$ and $[X]$ will be real symmetric. So the eigen pair $([I_m],\lambda_m)$ of \eqref{7} will be necessarily real valued. However it can be noted from \cite{sandip_temc1} that the lossy structures will have complex eigenvectors and eigenvalues. Following the orthogonal properties of the eigenvectors of \eqref{7}, it can be written that
  \begin{subequations}
\begin{align}
  [I_m]^H[R][I_n]=\delta_{mn}[I_m]^H[R][I_m] \label{8a}\\
  [I_m]^H[X][I_n]=\lambda_m\delta_{mn}[I_m]^H[R][I_m]\label{8b}\\
  \delta_{mn}=1 ~\text{if}~m=n,~\text{else}~0\label{8c}
\end{align}
\end{subequations}
The Hermitian transpose operation is denoted by \enquote{H}. for the lossless structures, it will turn out to be as transposition operation. The modal significance of the $m^{th}$ eigenmode is denoted by
\begin{equation}
   MS_m=\frac{1}{\sqrt{1+\lambda_m^2}}.
    \label{9}
\end{equation}   

Following the complex -power balance relation of \cite{harrington1971theory}, the power radiated by the $m^{th}$ mode can be defined as,
 \begin{equation}
   P_{R,m}=|\alpha_m|^2=MS_m (\frac{[I_m]^H[V]}{[I_m]^H[R][I_m] })^2
  \label{10}
\end{equation} 
Total radiated power at a certain frequency,say,$f_0$ will be a weighted sum of all individual modal power,
\begin{equation}
   P_{R}=\sum_{m=1}^{M}P_{R,m}=\sum_{m=1}^{M}|\alpha_m|^2
    \label{11}
\end{equation} 

Now, let's come back to the first two design objectives as stated in the initial part of the previous section. It can be noted from \eqref{10} that the outward radiated power $P_{R,m}$ is inversely related to the eigenvalue $\lambda_m$ of the corresponding mode.  As $\lambda_m$ tends towards 0, corresponding modal significance $ MS_m$ of \eqref{9} and radiated power $P_{R,m}$ of \eqref{11} will be higher. Similarly when the antenna is used in the receiving end, real power of the pointing vectors provides $P_{R,m}$ as the received real power. So, if the excitation vector is restricted to a very finite region compared to the total geometry of the antenna structure, the optimal frequency of maximum radiation can be approximated following the minima of the corresponding eigenvalue. 

\begin{figure}[ht!]%
\subfigure[]{%
\label{fig1a}%
\centering
\includegraphics[width=0.4\textwidth,clip]{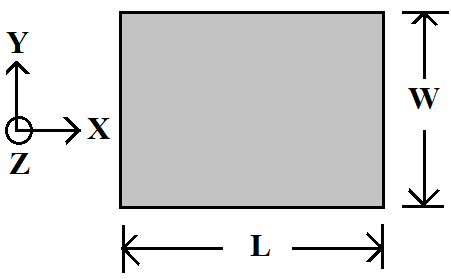}}
\hfill
\subfigure[]{%
\label{fig1b}%
\centering
\includegraphics[width=0.4\textwidth,clip]{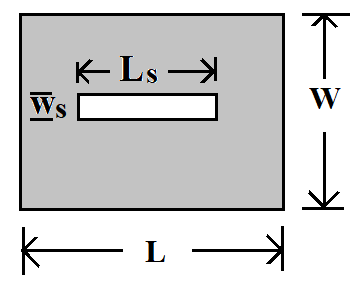}}
\caption{(a) Reference plate with $L=50$ mm, $W=40$ mm. (b)The plate with narrow slot. For X-directed slot, $L_s=30$ mm and $W_s=0.2$ mm. For Y-directed slot, $L_s=0.2$ mm and $W_s=30$ mm.}
\end{figure}

On the other hand, total radiated power consists of the contribution from all modes in \eqref{11}. To align the field distribution along a certain direction or to achieve a desired SAR distribution, it is required to suppress the adjacent undesired modes. Let the $i^{th}$ mode generates the desired beam pattern at the resonating frequency of $\omega_i$. Henceforth the desired radiated modal power is represented by $|\alpha_m|^2$ where $m=i$. Alternately, the undesired component of the radiated power are contributed by the other additional modes, $\sum_{m=1}^{M}|\alpha_m|^2$ for $m \neq i$.  So the control over undesired radiation needs to start from the search of the characteristic mode(/s) whose modal radiated power is comparable to the desired mode's power. Physically the eigenvalue parameter $\lambda_i$ denotes the proximity towards resonance of the $i^{th}$ characteristic mode. Ideal resonance criterion is defined by zero reactive power,in other words, $\lambda_m=0$. So, the dominant eigenmodes are commonly determined using the modal significance parameter $MS_m$ of \eqref{9}. The next discussion shows how the modal characteristics vary due to loading of narrow slot. For the purpose of illustration, let us consider a rectangular PEC plate in Fig.\ref{fig1a}. The plate loaded with narrow slot is shown in Fig.\ref{fig1b}.

\begin{figure}[ht!]%
\subfigure[Mode 1, Ref.]{%
\label{fig2a}%
\centering
\includegraphics[width=0.42\textwidth,clip]{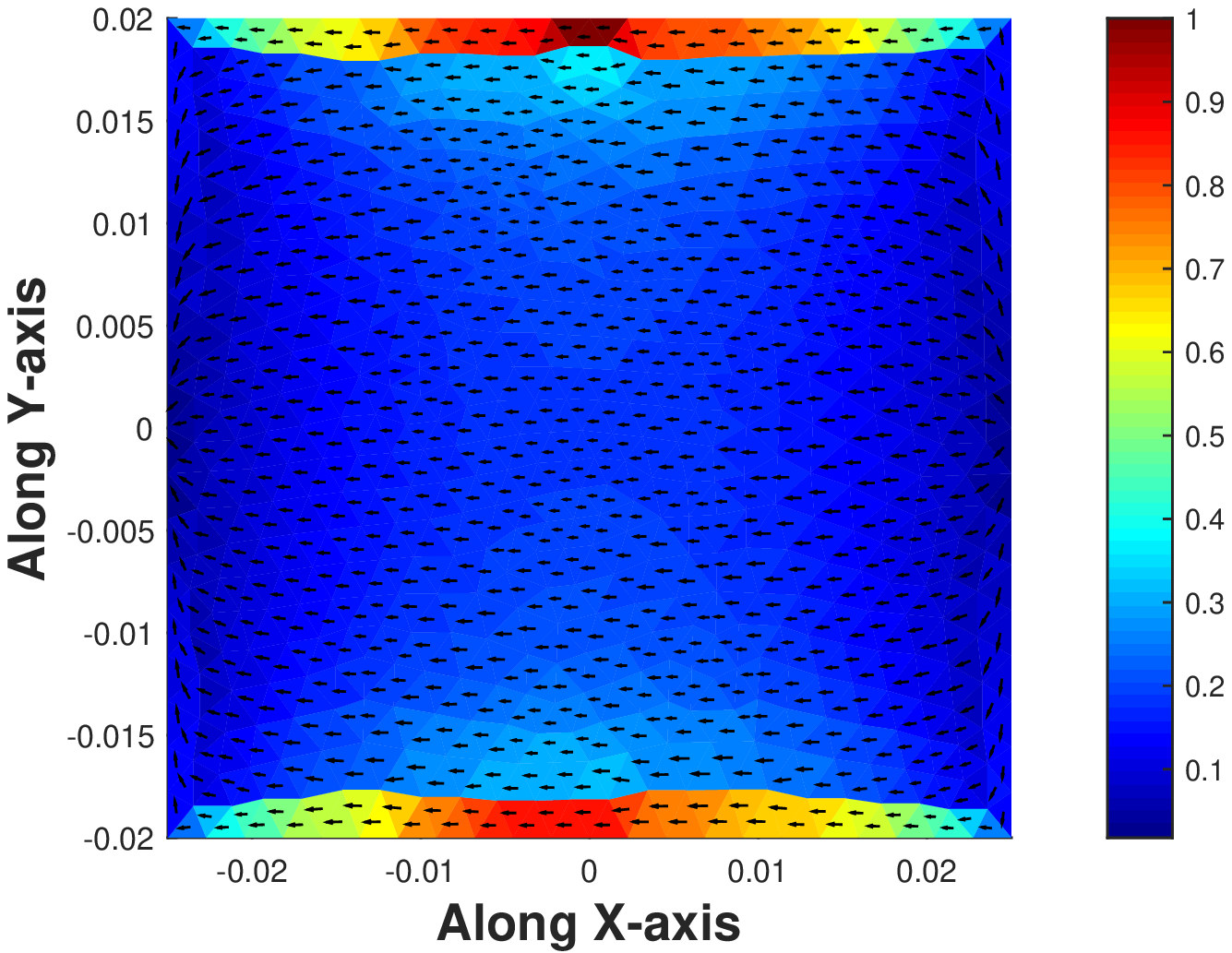}}
\hfill
\subfigure[Mode 2, Ref.]{%
\label{fig2b}%
\centering
\includegraphics[width=0.42\textwidth,clip]{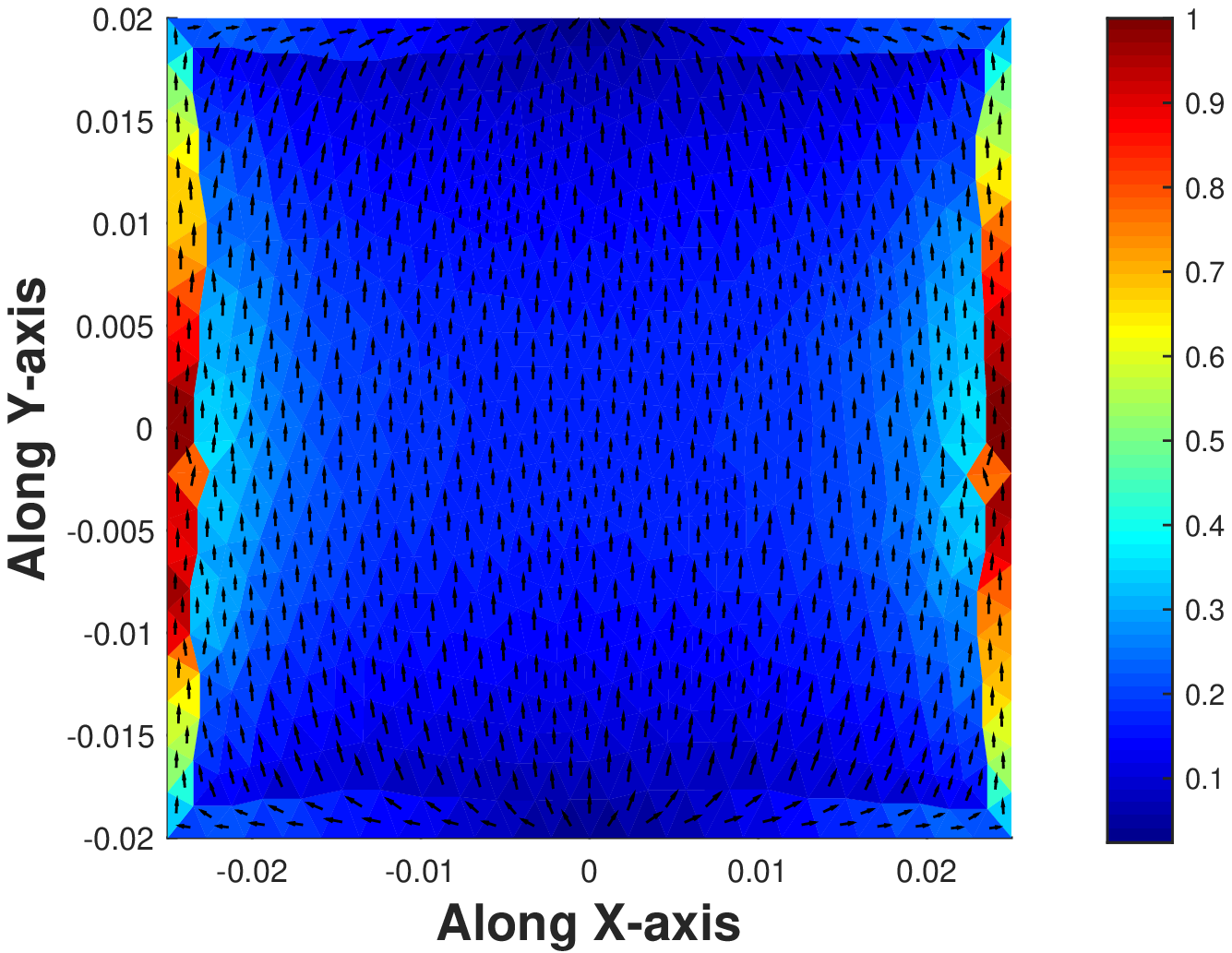}}
\hfill
\subfigure[Mode 1, X-slot]{%
\label{fig2c}%
\centering
\includegraphics[width=0.42\textwidth,clip]{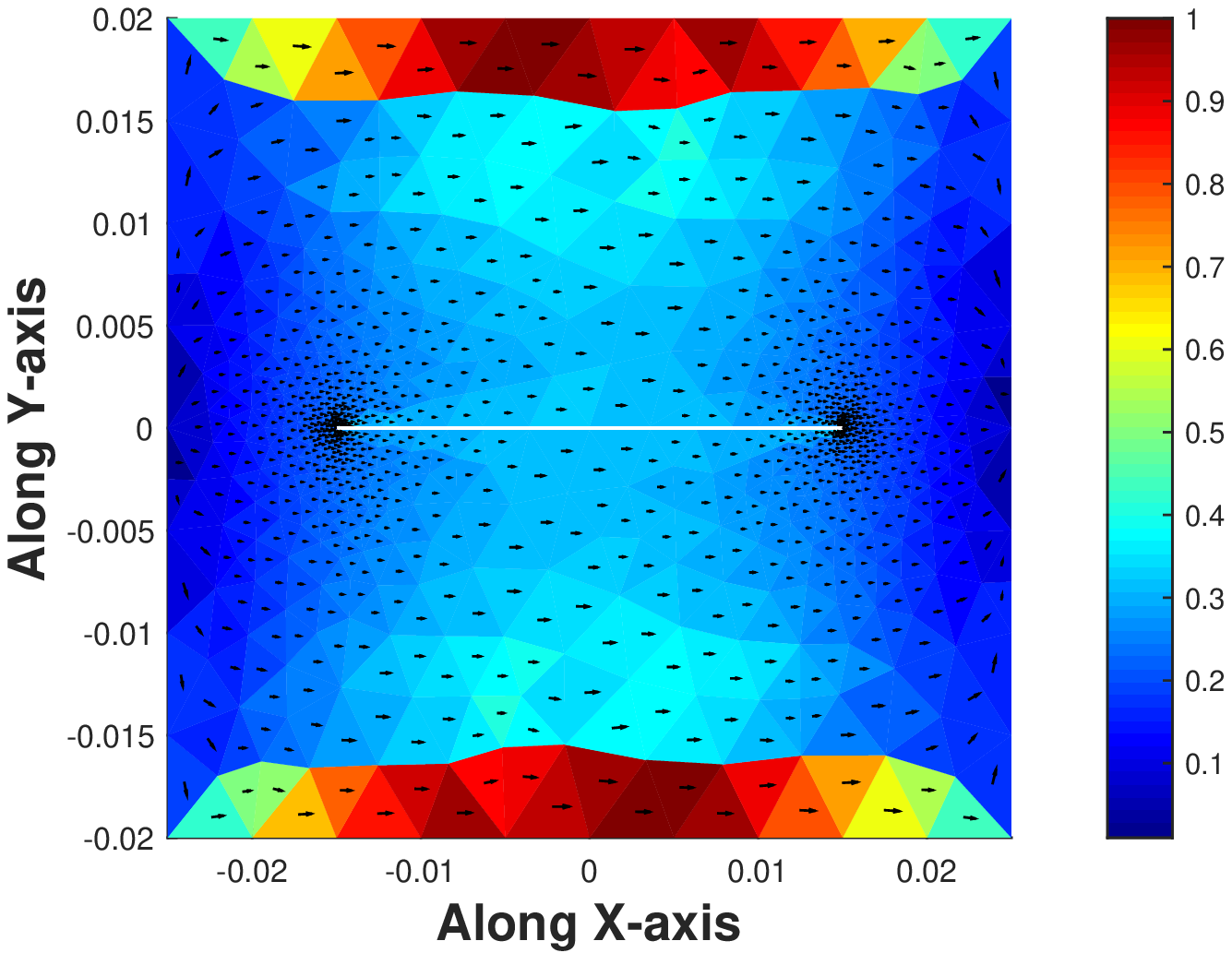}}
\hfill
\subfigure[Mode 2, X-slot]{%
\label{fig2d}%
\centering
\includegraphics[width=0.42\textwidth,clip]{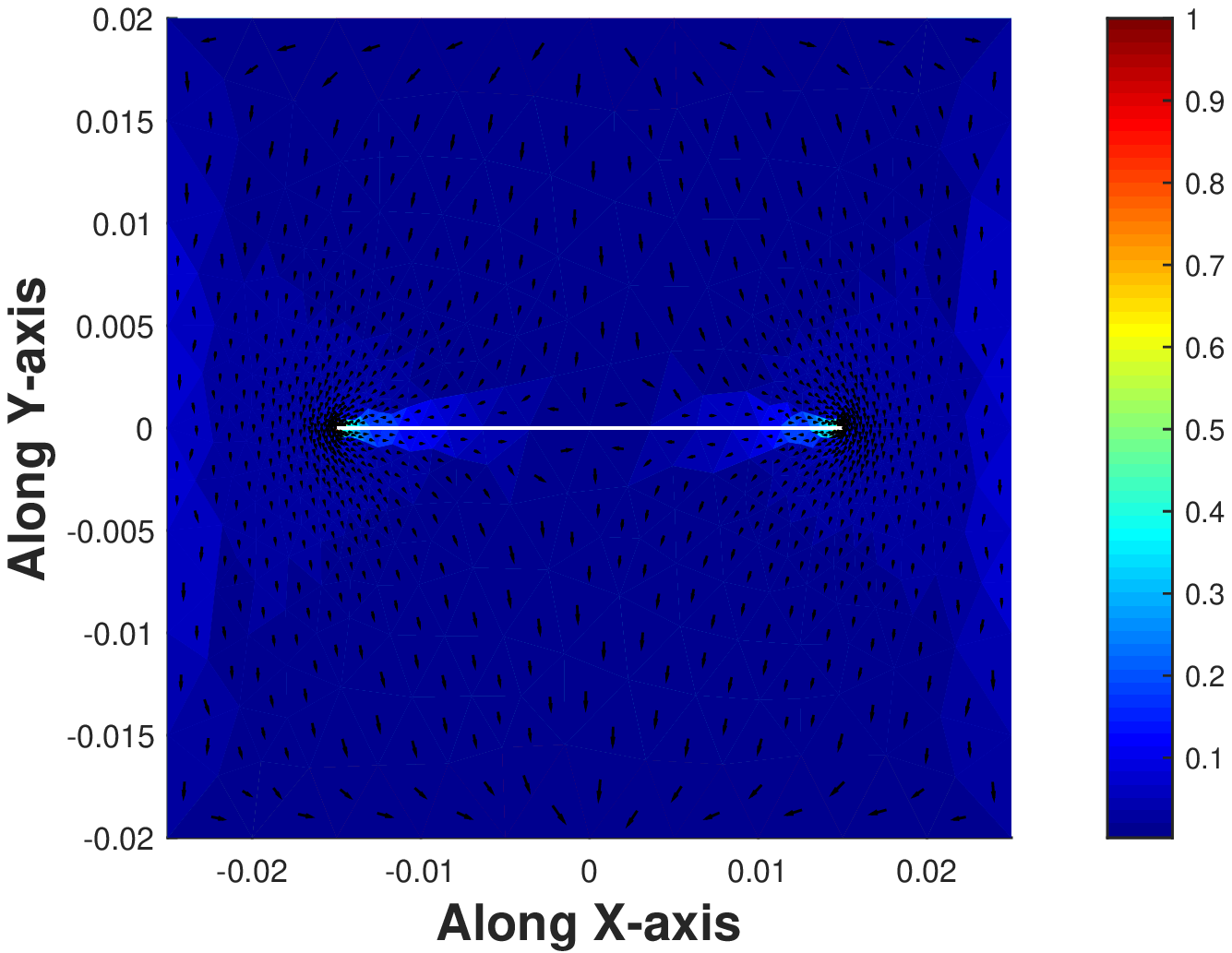}}
\hfill
\subfigure[Mode 1, Y-slot]{%
\label{fig2e}%
\centering
\includegraphics[width=0.42\textwidth,clip]{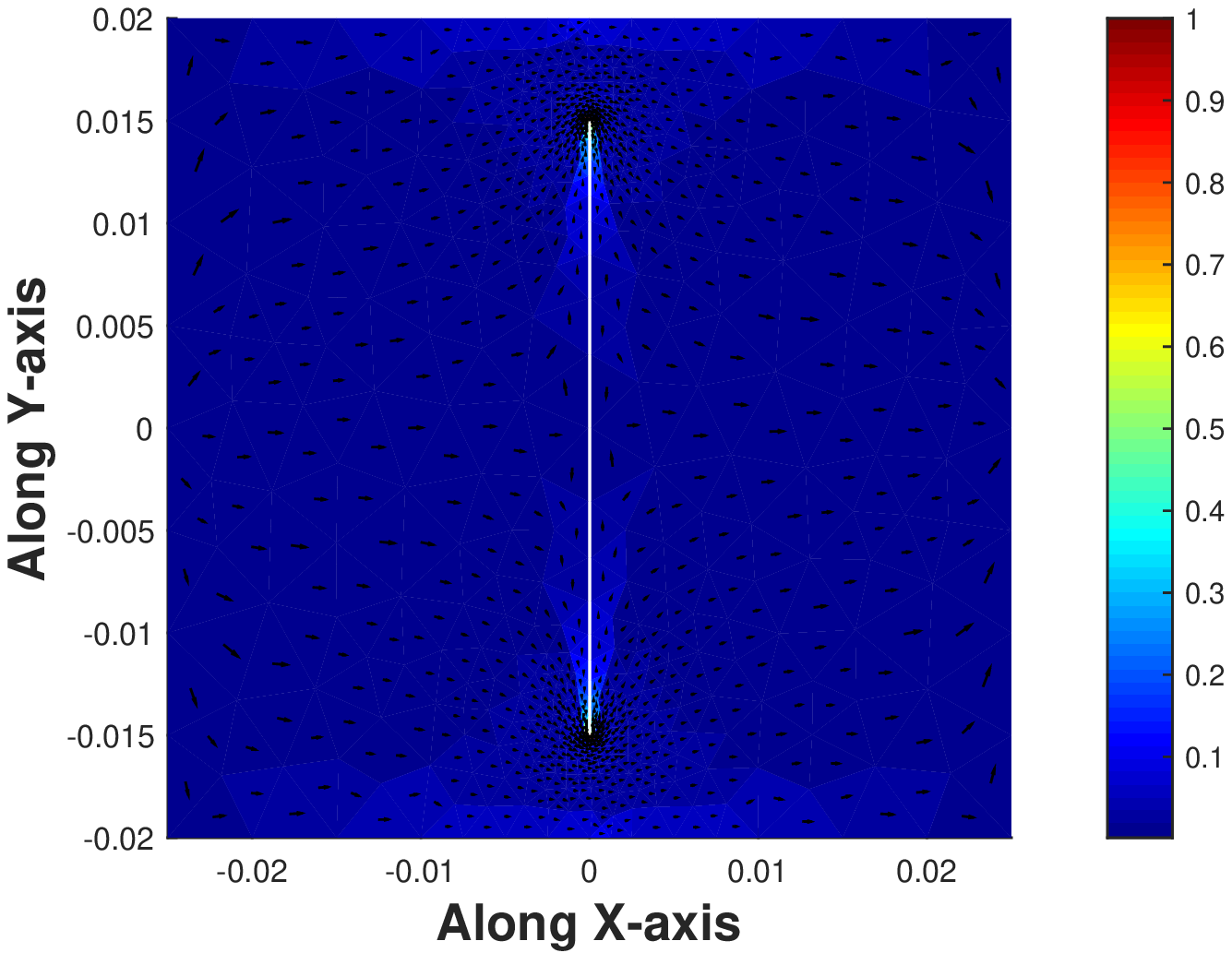}}
\hfill
\subfigure[Mode 2, Y-slot]{%
\label{fig2f}%
\centering
\includegraphics[width=0.42\textwidth,clip]{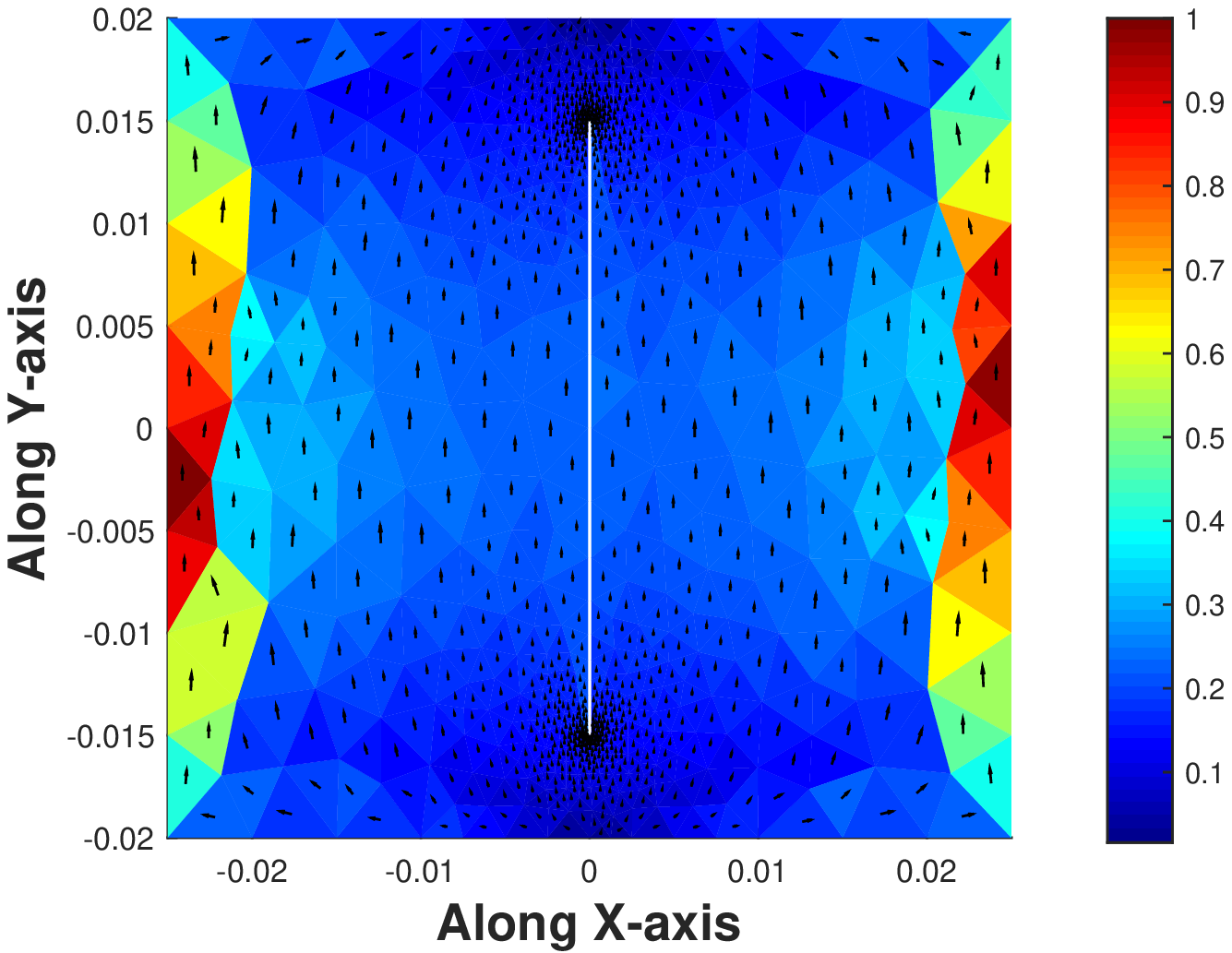}}
\caption{Normalized distribution of the $1^{st}$ and $2^{nd}$ eigencurrent modes at $f=2.4$ GHz.}
\end{figure}

The eigenmodes have been computed in MATLAB \cite{matlab} following \cite{harrington1971theory} and \cite{makarov2002antenna} with RWG basis functions. Sorting the modes following \eqref{9}, the $1^{st}$ two dominant eigen current modes at $f=2.4$ GHz have been plotted in Figs.\ref{fig2a} and \ref{fig2b}. Keeping both the current distribution in same scale, the eigencurrent modes have been normalised with respect to its maximum amplitude. This leads to the presence of the additional term $[I_m]^H[R][I_m]$ in defining $\alpha_m$ of \eqref{5}. As shown in  Figs.\ref{fig2a} and \ref{fig2b}, the $1^{st}$ and $2^{nd}$ eigenmodes are found to be linearly polarized along X and Y-direction respectively.\\
In the next stage, narrow slot has been incorporated with its wider axis along the X-direction where $Ws=0.2$ mm and $Ls=30$ mm. The perturbed current distribution of the first two dominant modes at $f=2.4$ GHz are shown in Figs. \ref{fig2c} and \ref{fig2d}. Major contribution of the slot is reflected in the second mode of Fig. \ref{fig2d}. It can be noted from the comparison between Fig.\ref{fig2b} and \ref{fig2d} that the X-directed slot converts the nearest Y-polarized edge mode of Fig.\ref{fig2b} into a slot mode of Fig.\ref{fig2d} where the current distribution hold significant magnitude only along two narrow ends of the slot. The X-directed slot has minimal impact on the first mode in Fig.\ref{fig2c} whose orientation is parallel to the wider axis of the slot. For studying the complementary case, the slot has been aligned along the Y-direction and the resultant modes have been shown in Figs. \ref{fig2e} and \ref{fig2f}. Similar to the earlier case, the Y-polarised edge mode of Fig.\ref{fig2a} is transformed into the Y-polarised edge mode in Fig.\ref{fig2e}. Minimal variation is noted in the X-polarised edge mode of Fig.\ref{fig2f}.

\begin{figure}[ht!]
\centering
\includegraphics[width=1.8in]{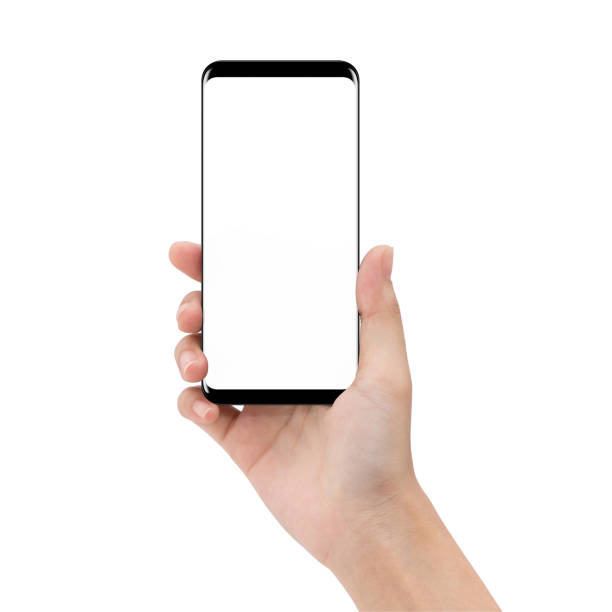}
\label{figg3}
\caption{Field observation plane at $z=5$ mm height from the radiating plate.}
\end{figure}

\begin{figure}[ht!]%
\subfigure[$20log_{10}|\vec{E}_1|$ along X-axis]{%
\label{fig4a}%
\centering
\includegraphics[width=0.45\textwidth,clip]{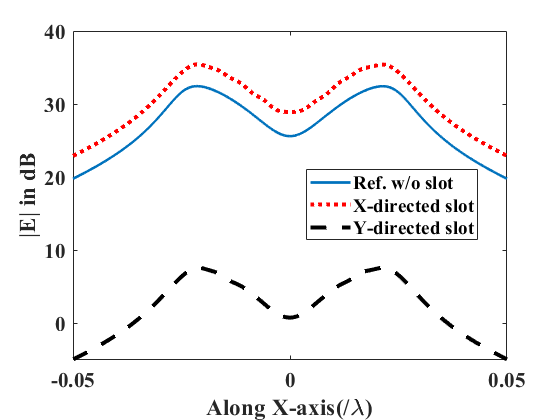}}
\hfill
\subfigure[$20log_{10}|\vec{E}_1|$ along Y-axis]{%
\label{fig4b}%
\centering
\includegraphics[width=0.45\textwidth,clip]{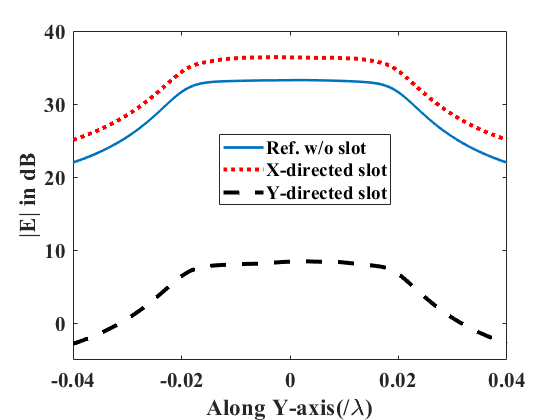}}
\hfill
\subfigure[$20log_{10}|\vec{E}_2|$ along X-axis]{%
\label{fig4c}%
\centering
\includegraphics[width=0.45\textwidth,clip]{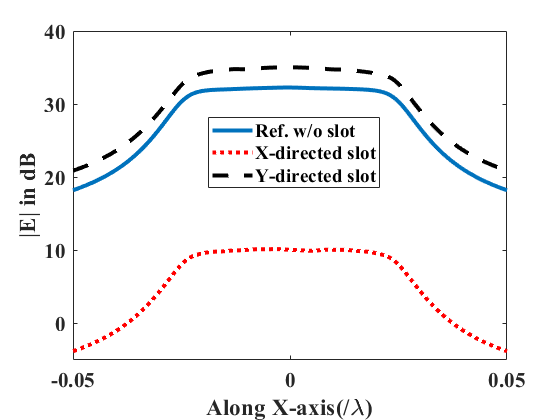}}
\hfill
\subfigure[$20log_{10}|\vec{E}_2|$ along X-axis]{%
\label{fig4d}%
\centering
\includegraphics[width=0.45\textwidth,clip]{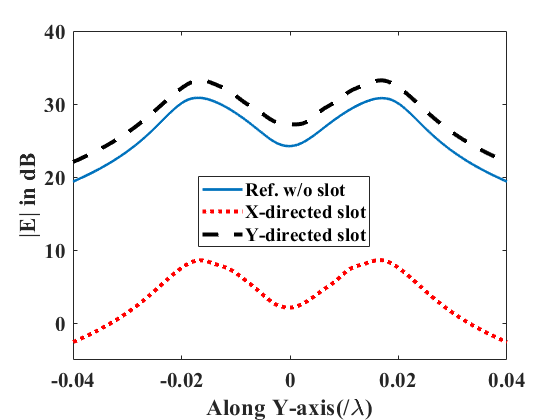}}
\caption{Near field distribution due to the first two dominant characteristic current modes at $f=2.4$ GHz.}
\end{figure}

As the loading of slot modify the current distribution, corresponding field distribution will also change. The SAR distribution on a nearby object is directly related to the electric field as
\begin{align}
\text{SAR}=\frac{\sigma}{\rho}|\text{E}|^2
\label{eq12}
\end{align}

The nearby object will experience the incident field $\text{E}$ contributed by the antenna structure. For example while holding a cell phone like Fig. \ref{figg3}, the horizontal radiation will influence the biological tissues. To understand the impacts on the near field variation along the horizontal direction, let us consider a line of observation at an offset distance from the edges of the PEC plate (i.e., the radiating source). Using infinitesimal dipole model of \cite{makarov2002antenna}, the electric field $20log_{10}|\text{E}|$ has been calculated for the corresponding eigencurrent modes. It can be noted from Figs. \ref{fig4a} and \ref{fig4b} that the X-directed slot enhances the X-polarised $1^{st}$ mode's near field density where as the Y-directed slot reduces the corresponding modal strength. Similarly, the Y-polarised mode is found to be weakened by the X-directed narrow slot in Figs. \ref{fig4c} and \ref{fig4d}. Such variation in the near field behaviour due to loading of narrow slot will influence the SAR distribution of the neighbouring biological tissue following \eqref{eq12}. Therefore, the radiating element can be suitably designed based the characteristic mode analysis to minimize the undesired radiation. Till now, the modal analysis does not consider the information of the excitation vector.
The next section uses the same radiating elements of Figs. \ref{fig1a} and \ref{fig1b} with finite microstip ground plane to verify the suitability of the slot loading method using full-wave simulator Ansys HFSS \cite{ansys2015ansoft}.


\begin{figure}[ht!]
\centering
\includegraphics[width=2.5in]{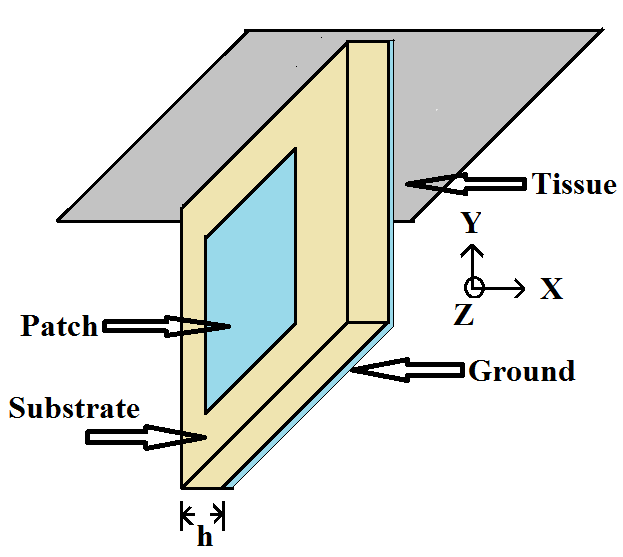}
\caption{Antenna with the adjacent tissue layer.}
\label{fig:tissue}
\end{figure} 
\newpage
\section{Full-wave Analysis}
For the full-wave study, two different case studies have been considered with probe-feeding and microstrip line feeding arrangement. For each case, the patch dimensions have been assumed to be same as of Figs. \ref{fig1a} and \ref{fig1b}. The patch surface is lying 1.57 mm above the lossless Roger substrate with the dielectric permittivity of 2.2. The tissue is placed parallel to the XZ-plane with the Y-separation of 2 mm from the substrate edge of the antenna as shown in Fig. \ref{fig:tissue}.
\subsection{Probe-fed Patch}
\begin{figure}[ht!]%
\subfigure[]{%
\label{fig6a}%
\centering
\includegraphics[width=0.45\textwidth,clip]{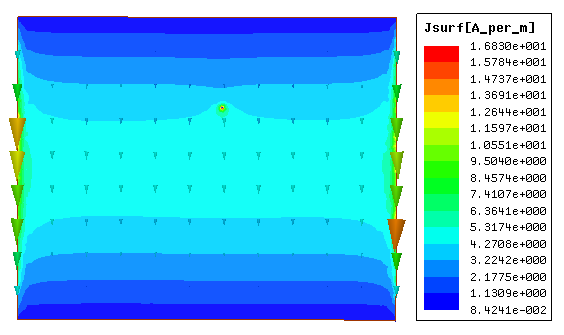}}
\hfill
\subfigure[]{%
\label{fig6b}%
\centering
\includegraphics[width=0.45\textwidth,clip]{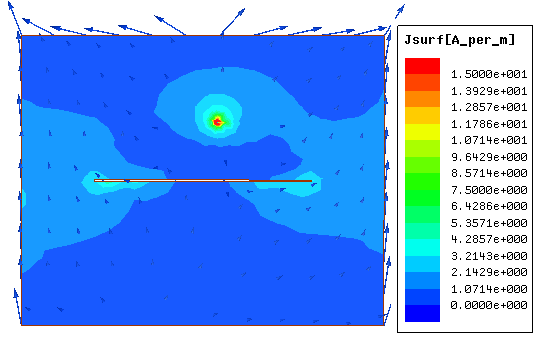}}
\hfill
\subfigure[]{%
\label{fig6c}%
\centering
\includegraphics[width=0.45\textwidth,clip]{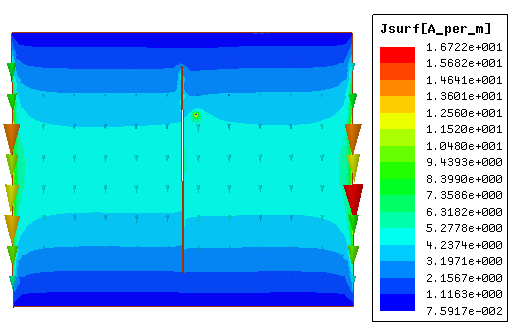}}
\hfill
\subfigure[]{%
\label{fig6d}%
\centering
\includegraphics[width=0.45\textwidth,clip]{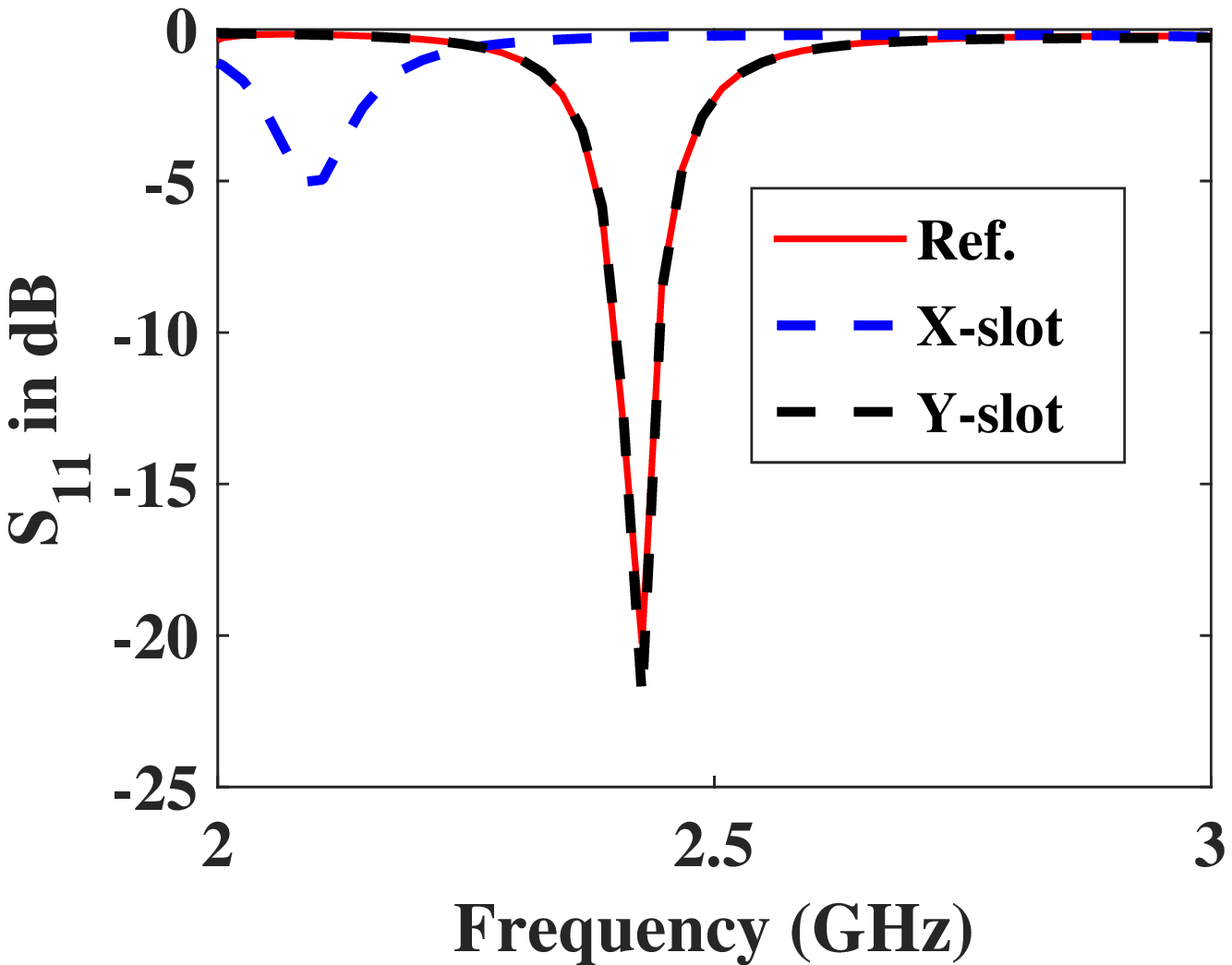}}
\caption{Normalized distribution of the characteristic current modes at $f=2.4$ GHz. (c) $1^{st}$ and (d) $2^{nd}$ modes of the plate with X-directed slot.}
\end{figure}

The dimension of the ground plane is assumed to be 83.6 mm $\times$ 71.5 mm for tuning the reference antenna around 2.4 GHz using Ansys Design Kit. The antenna has been excited by co-axial probe feeding at ($x_f=2$ mm, $y_f=8.05$ mm, $z_f=1.57$ mm). The current distribution induced on the patch surface of the reference antenna is shown in Fig. \ref{fig6a} where the the current is found to be Y-polarised. Following the previous modal discussion, if it is desired to suppress this Y-polarised current density, the narrow slot needs to be aligned along the orthogonal direction, i.e., along the X-axis. So for the sake of comparison, two examples have been studied with the X-directed and Y-directed slots on the patch surface. The feed location has been taken as same of the reference probe-fed patch. As obvious in Fig. \ref{fig6a}, the orthogonal orientation of the slot can perturb the Y-polarised edge mode of Fig.  \ref{fig6a} into the slot mode. On the other hand, as the slot is oriented parallel to the Y-polarised edge mode, it has negligible impact on the current distribution in Fig. \ref{fig6c}. As the slot perturbs the induced current density vector, the impedance characteristics also change in Fig. \ref{fig6d}. The change in the equivalent circuit parameters can be computed using the existing analysis of \cite{ghosal2013analysis} and \cite{SG2018APS}. The next discussion shows how the perturbation of the current vector becomes reflected in the near field and  and corresponding SAR distribution. 

\begin{figure}[ht!]%
\subfigure[$20log_{10}|\vec{E}_1|$ along X-axis]{%
\label{figp4a}%
\centering
\includegraphics[width=0.45\textwidth,clip]{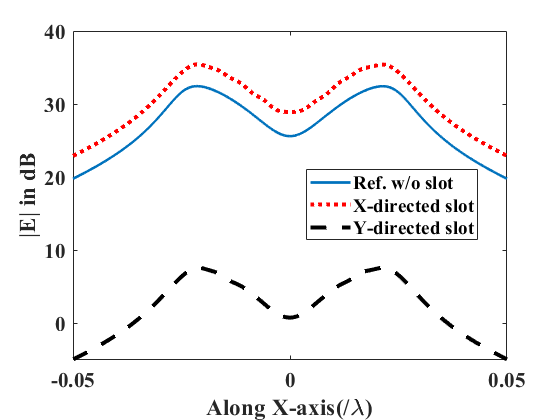}}
\hfill
\subfigure[$20log_{10}|\vec{E}_1|$ along Y-axis]{%
\label{figp4b}%
\centering
\includegraphics[width=0.45\textwidth,clip]{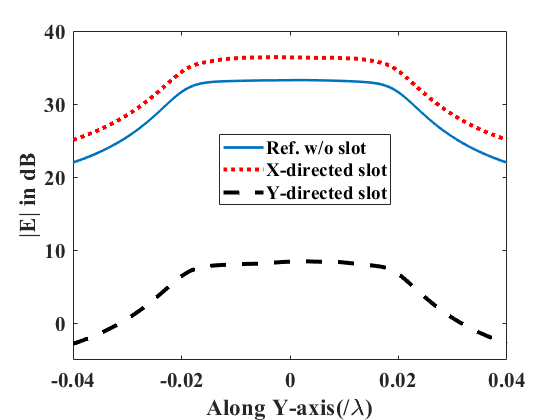}}
\caption{Near field distribution due to the probe-fed patch at $f=2.4$ GHz.}
\end{figure}

\begin{figure}[ht!]%
\subfigure[$\phi=0^{\circ}$-plane]{%
\label{figpf4a}%
\centering
\includegraphics[width=0.45\textwidth,clip]{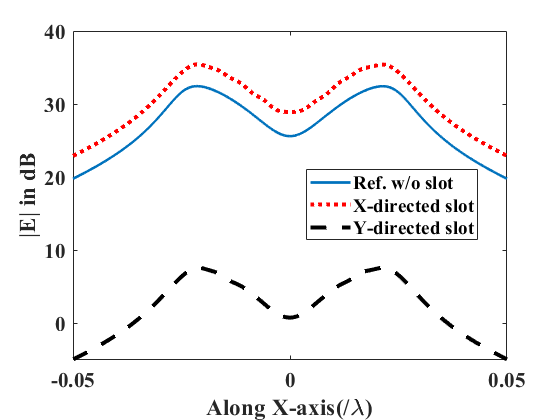}}
\hfill
\subfigure[$\phi=0^{\circ}$-plane]{%
\label{figpf4b}%
\centering
\includegraphics[width=0.45\textwidth,clip]{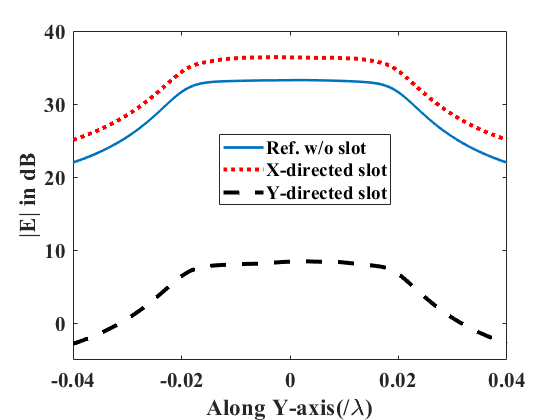}}
\caption{Far field distribution due to the probe-fed patch at $f=2.4$ GHz.}
\end{figure}

To understand the change in horizontal radiation, the near field at the same offset distance from two edges are reported in Figs. \ref{figp4a} and \ref{figp4b}. Similar to the previous observation of Figs. \ref{fig4a}--\ref{fig4d}, the near field strength along the horizontal directed is found to be reduced due to loading of the narrow slot of orthogonal orientation. In the next stage, the tissue layer is placed parallel to the Y-edge of the antenna as shown in Fig. \ref{fig:tissue}. The separation distance between the antenna and the tissue layer along the Y-axis is assumed to be 2 mm.

\begin{table}[ht!]
\begin{center}
\caption{Specification of the tissue layers }
\label{Table1}
\setlength{\extrarowheight}{1.7pt}
\scalebox{1.3}{
\begin{tabular}{|c|c|c|c|c|}
 \hline
Tissue & $\mathbf{\sigma(S/m)}$ & $\mathbf{\epsilon_r}$ & $\mathbf{\rho(kg/m^{3})}$ & d(mm)\\
 \hline
 skin & $1.43$ &$38.1$ &1100 &1.5\\
 \hline
 fat & $0.1$ &$5.29$ &916 &1.5\\
 \hline
 \end{tabular}}
\end{center}
\end{table}

Following \cite{zhu2016miniaturized}, the material properties of the tissue model have been given in Table \ref{Table1} where the thickness of the corresponding layer is denoted by $d$ in mm. The electrical conductivity, relative permittivity and mass density of the layers are denoted by $\rho$, $\epsilon_r$ and $\sigma$ in Table \ref{Table1}. Since the tissue layer is placed with very close electrical distance ($0.016 \lambda_0$, $\lambda_0$ being the free-space wavelength at $f=2.4$ GHz) from the antenna element, it lies in the radiative near field region at $f=2.4$ GHz. 
\begin{figure}[ht!]%
\subfigure[Ref.,skin]{%
\label{fig7a}%
\centering
\includegraphics[width=0.45\textwidth,clip]{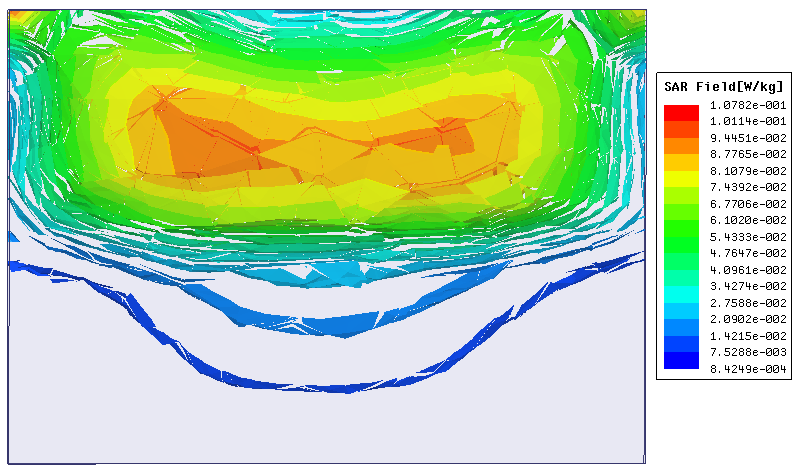}}
\subfigure[Ref.,fat]{%
\label{fig7b}%
\centering
\includegraphics[width=0.45\textwidth,clip]{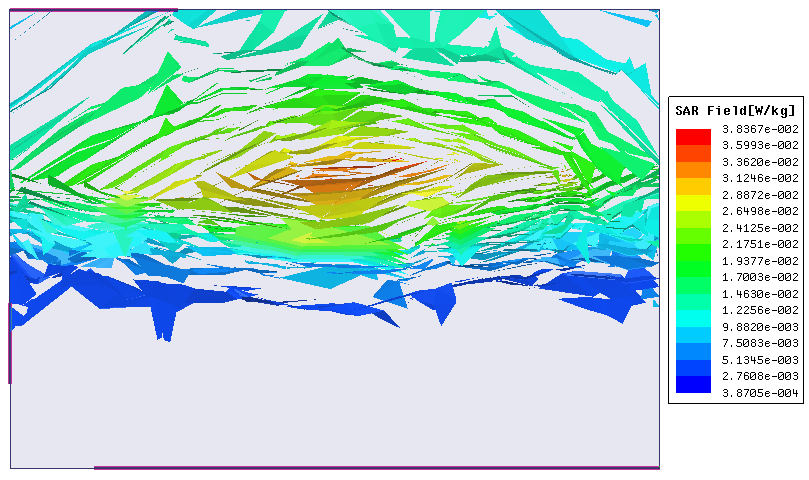}}
\hfill
\subfigure[X-slot,skin]{%
\label{fig7c}%
\centering
\includegraphics[width=0.45\textwidth,clip]{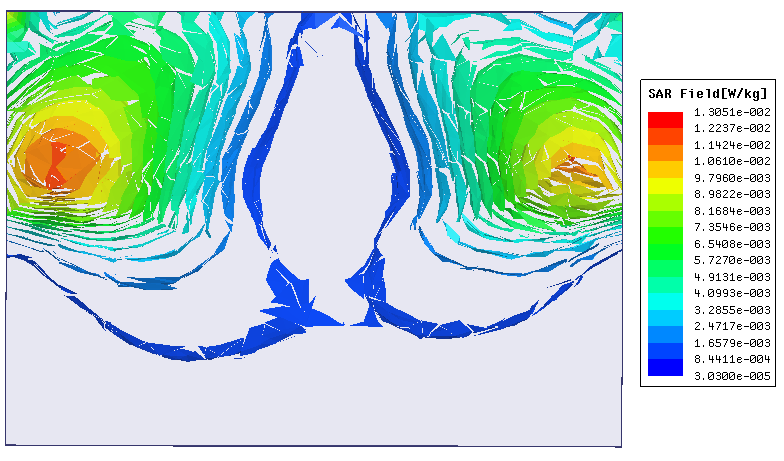}}
\hfill
\subfigure[X-slot,fat]{%
\label{fig7d}%
\centering
\includegraphics[width=0.45\textwidth,clip]{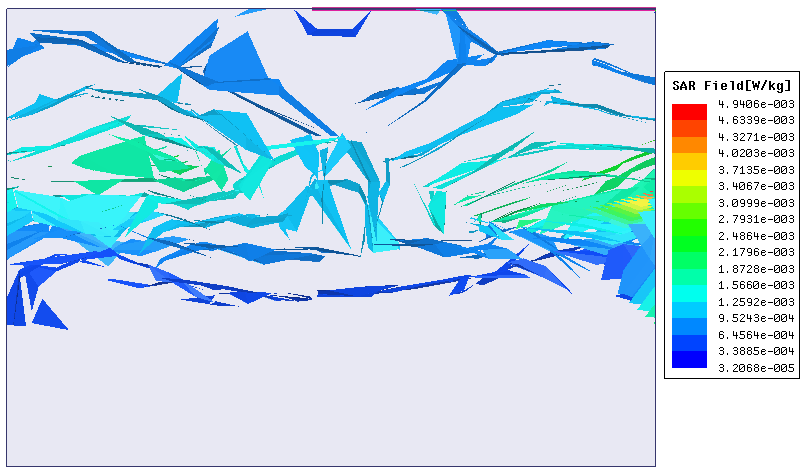}}
\hfill
\subfigure[Y-slot,skin]{%
\label{fig7e}%
\centering
\includegraphics[width=0.45\textwidth,clip]{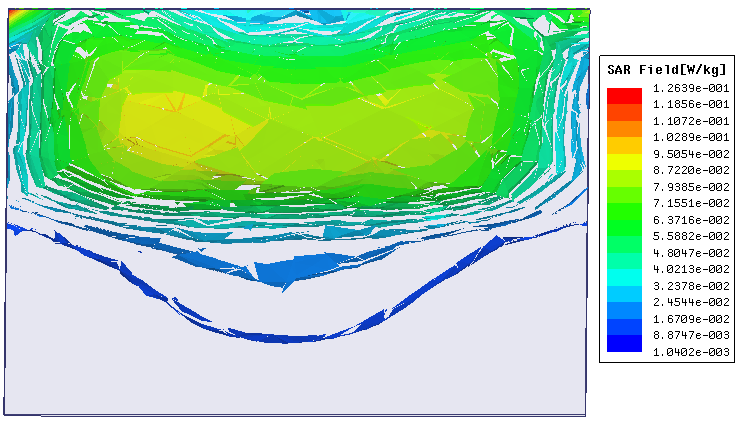}}
\hfill
\subfigure[Y-slot,fat]{%
\label{fig7f}%
\centering
\includegraphics[width=0.45\textwidth,clip]{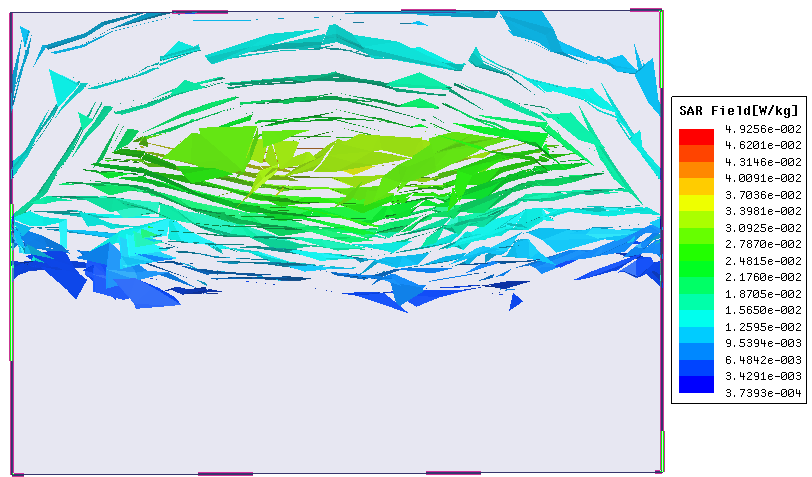}}
\caption{SAR distribution for three probe-fed antenna topologies with two different tissue layers at $f=2.4$ GHz. Tissue is lying along the XZ-plane with the dimension of $85$ mm $\times 60$ mm.}
\end{figure}

Being  placed at horizontal separation, the tissue will be coupled by the horizontal radiation components. The electromagnetic exposure of the tissue layer can be characterised by specific absorption rate (SAR) parameter of \eqref{eq12}. The incident field on the tissue is directly related to the induced characteristic modes of the antenna which vary its nature due to orientation of the narrow slot. The field emerging out of the radiating patch antenna will experience multiple reflection and scattering from the tissue layer. As a consequence, the SAR distribution on the tissue will be perturbed accordingly. Apart from the induced field $\vec{E}_
o$, effective amount of the SAR value is also dependent on $\rho$ and $\sigma$. For the case study, two different tissues-skin and fat have been considered.  The SAR distribution for the probe-fed reference patch have been computed using Ansys HFSS and shown in Figs. \ref{fig7a}--\ref{fig7f}. It can be noted from the comparison of Figs. \ref{fig7a}, \ref{fig7c}, and \ref{fig7e} that the X-directed slot reduces the electromagnetic energy absorption compared to the reference antenna with no slot. The reason will be obvious from the preceding current distribution and near filed distribution where the Y-directed slot suppresses the Y-polarised current distribution of the reference patch.
\begin{table}[ht!]
\begin{center}
\caption{Maximum SAR value comparison of the probe-fed patch }
\label{Table2}
\setlength{\extrarowheight}{1.7pt}
\scalebox{1.3}{
\begin{tabular}{|c|c|c|}
 \hline
Tissue & $\frac{max(SAR_{ref})}{max(SAR_{X-slot})}$ & $\frac{max(SAR_{ref})}{max(SAR_{Y-slot})}$\\
 \hline
 skin & $8.23$ &$0.8492$\\
 \hline
 fat & $7.77$ &$0.78$\\
 \hline
 \end{tabular}}
\end{center}
\end{table}

Alternately, the Y-directed slot enhances the Y-polarised mode. As a consequence, the SAR value is found to be slightly increased for the patch with the Y-directed slot. The slot mode nature leads to generate two concentric maxima regions in the skin layer of Fig. \ref{fig7c}. Similar characteristics can be found the examples with fat layer also. The SAR value is reduced by the X-directed slot and enhanced by the Y-directed slot.  Maximum SAR values for the tissue layers have been compared in Table \ref{Table2} for different orientation of the slot. The next discussion considers microstrip line fed patches for comparison.

\begin{figure}[ht!]%
\subfigure[]{%
\label{fig8a}%
\centering
\includegraphics[width=0.45\textwidth,clip]{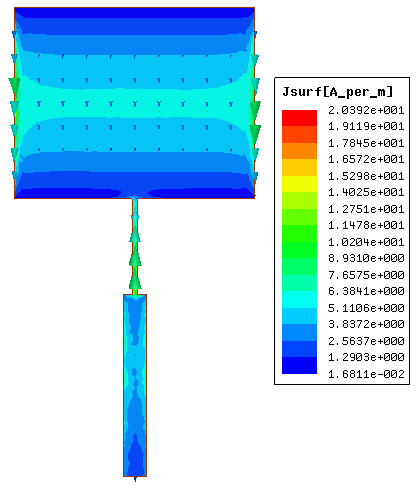}}
\hfill
\subfigure[]{%
\label{fig8b}%
\centering
\includegraphics[width=0.45\textwidth,clip]{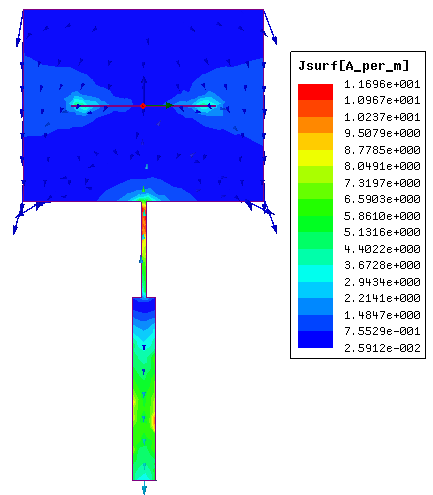}}
\hfill
\subfigure[]{%
\label{fig8c}%
\centering
\includegraphics[width=0.45\textwidth,clip]{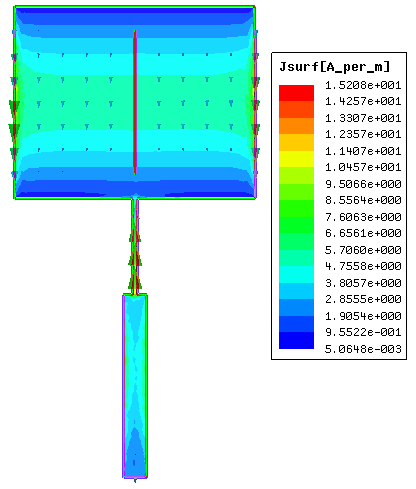}}
\hfill
\subfigure[]{%
\label{fig8d}%
\centering
\includegraphics[width=0.45\textwidth,clip]{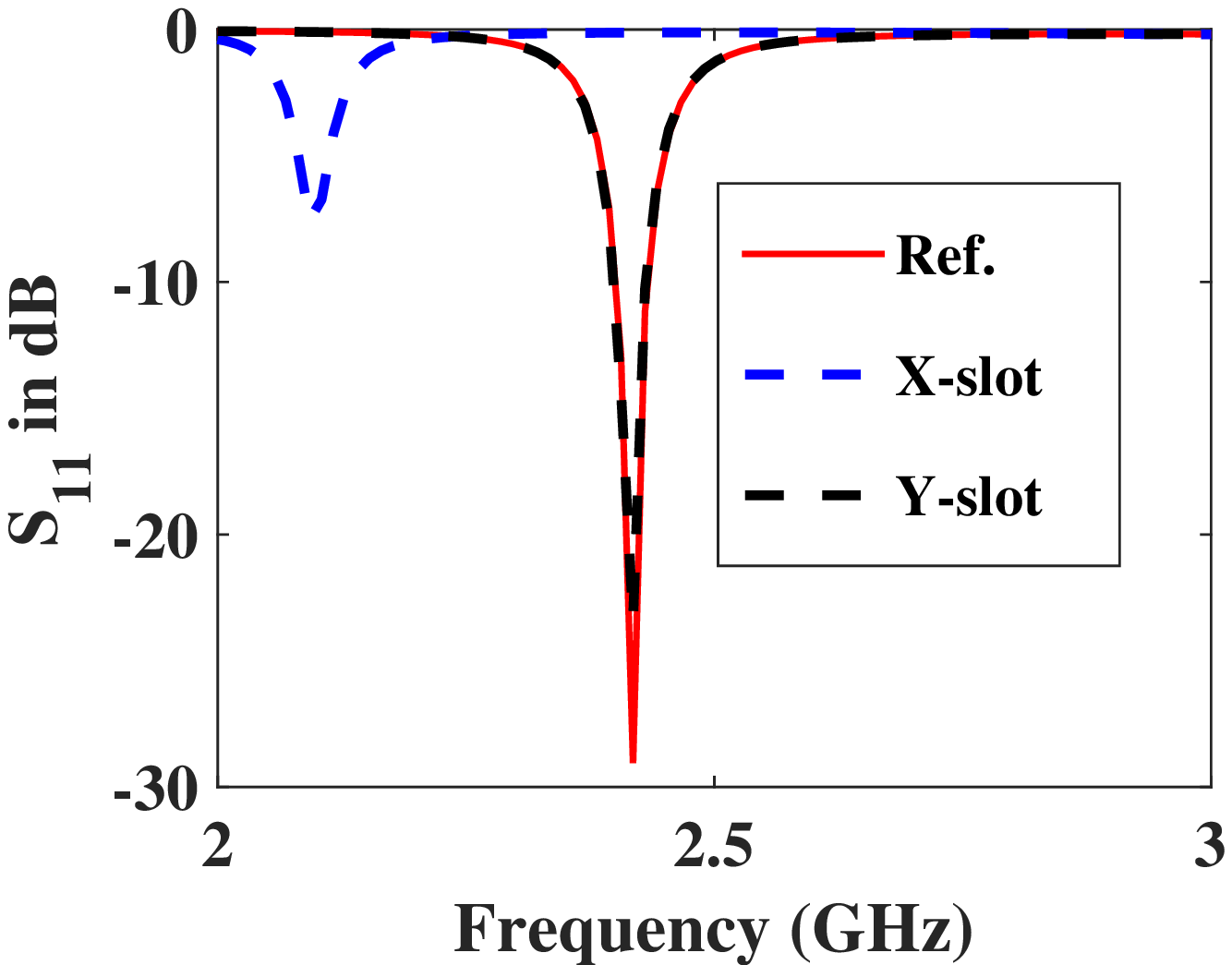}}
\caption{Normalized distribution of the characteristic current modes at $f=2.4$ GHz. (c) $1^{st}$ and (d) $2^{nd}$ modes of the plate with X-directed slot.}
\end{figure}

\subsection{Microstrip line fed Patch}
Similar patch elements have been considered but with microstrip line feeding arrangement. Using Ansys Design Kit, the feed-line of the reference antenna is tuned to 2.4 GHz with the same dimension of the radiating patch. The dimension of the ground plane is considered to be 83.6 mm $\times$ 156 mm. The current distribution have been compared in Figs. \ref{fig8a}--\ref{fig8c}. Similar to the probe-fed examples, the X-directed slot is found to suppress the Y-polarised edge mode of Fig. \ref{fig8a}. The impedance matching behaviour is also influenced by the orientation of the slot in Fig. \ref{fig8d}. \begin{figure}[ht!]
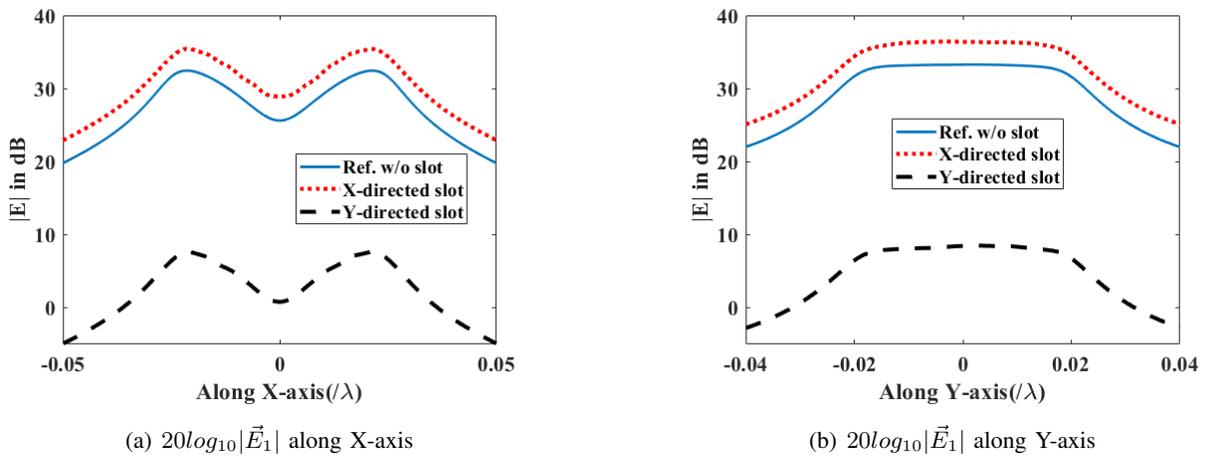
%
\subfigure[$20log_{10}|\vec{E}_1|$ along X-axis]{%
\label{figm4a}%
\centering
\includegraphics[width=0.45\textwidth,clip]{fig/E1_alongX.png}}
\hfill
\subfigure[$20log_{10}|\vec{E}_1|$ along Y-axis]{%
\label{figm4b}%
\centering
\includegraphics[width=0.45\textwidth,clip]{fig/E1_alongY.png}}
\caption{Near field distribution due to the microstrip line-fed patch at $f=2.4$ GHz.}
\end{figure}
\begin{figure}[ht!]%
\subfigure[$\phi=0^{\circ}$-plane]{%
\label{figmf4a}%
\centering
\includegraphics[width=0.45\textwidth,clip]{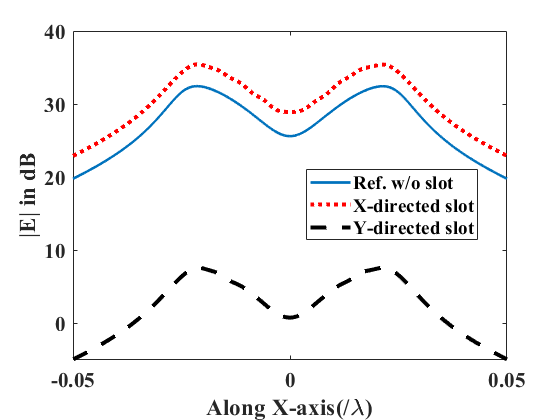}}
\hfill
\subfigure[$\phi=0^{\circ}$-plane]{%
\label{figmf4b}%
\centering
\includegraphics[width=0.45\textwidth,clip]{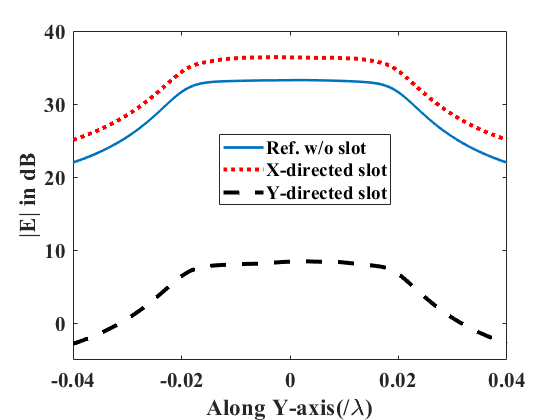}}
\caption{Far field distribution due to the microstrip line-fed patch at $f=2.4$ GHz.}
\end{figure}

\begin{figure}[ht!]%
\subfigure[Ref.,skin]{%
\label{fig9a}%
\centering
\includegraphics[width=0.45\textwidth,clip]{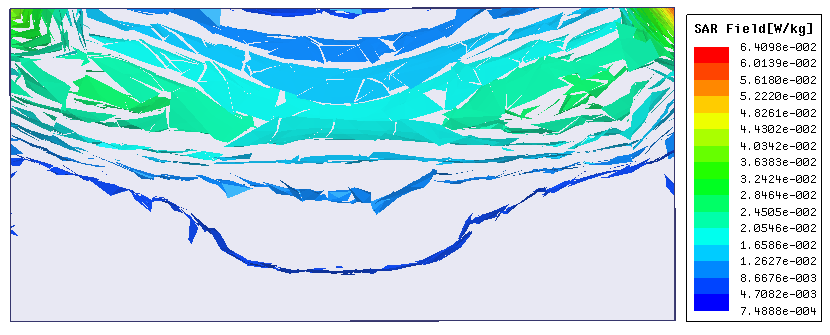}}
\subfigure[Ref.,fat]{%
\label{fig9b}%
\centering
\includegraphics[width=0.45\textwidth,clip]{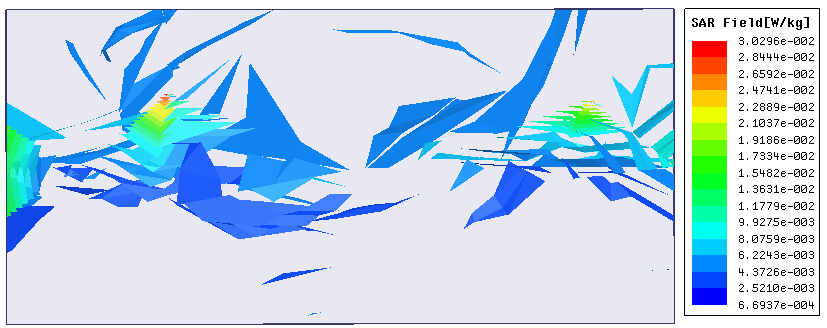}}
\hfill
\subfigure[X-slot,skin]{%
\label{fig9c}%
\centering
\includegraphics[width=0.45\textwidth,clip]{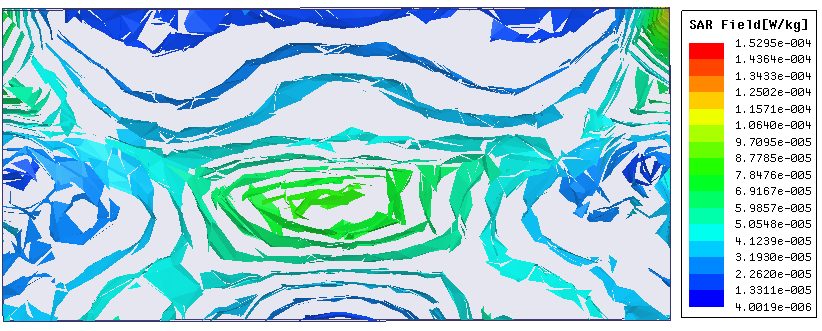}}
\hfill
\subfigure[X-slot,fat]{%
\label{fig9d}%
\centering
\includegraphics[width=0.45\textwidth,clip]{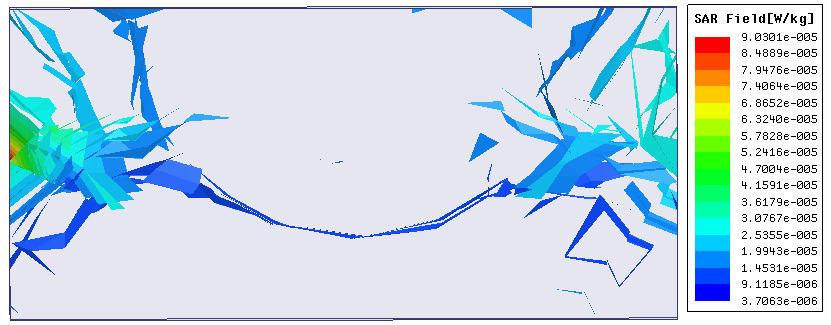}}
\hfill
\subfigure[Y-slot,skin]{%
\label{fig9e}%
\centering
\includegraphics[width=0.45\textwidth,clip]{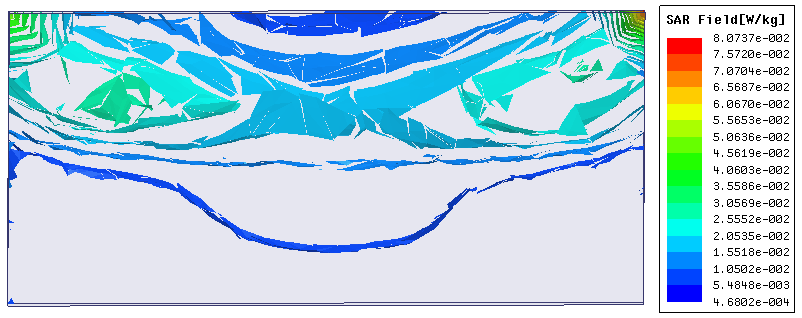}}
\hfill
\subfigure[Y-slot,fat]{%
\label{fig9f}%
\centering
\includegraphics[width=0.45\textwidth,clip]{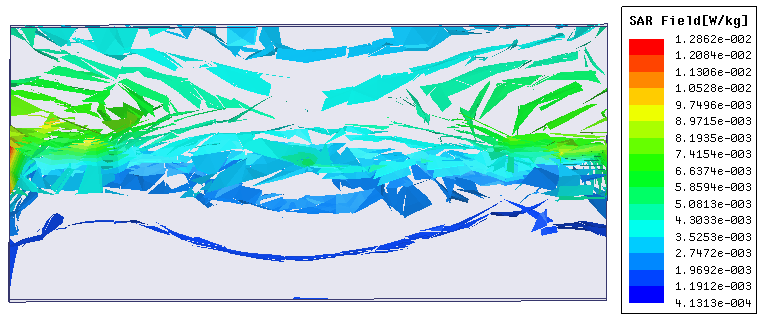}}
\caption{SAR distribution for three line-fed antenna topologies with two different tissue layers at $f=2.4$ GHz. Tissue is lying along the XZ-plane with the dimension of $85$ mm $\times 40$ mm.}
\end{figure}

As shown in Figs. \ref{figm4a} and \ref{figm4b}, the near field due to the horizontal radiation is also decreased by the orthogonal slot. Henceforth, the intuitive hints provided by the characteristic mode analysis is valid for both types of feeding arrangements.

\begin{table}[ht!]
\begin{center}
\caption{Maximum SAR value comparison of the line-fed patch }
\label{Table3}
\setlength{\extrarowheight}{1.7pt}
\scalebox{1.3}{
\begin{tabular}{|c|c|c|}
 \hline
Tissue & $\frac{max(SAR_{ref})}{max(SAR_{X-slot})}$ & $\frac{max(SAR_{ref})}{max(SAR_{Y-slot})}$\\
 \hline
 skin & $418.3$ &$0.79$\\
 \hline
 fat & $336.56$ &$2.36$\\
 \hline
 \end{tabular}}
\end{center}
\end{table}

The impacts of the current perturbation on the resultant SAR distribution have been studied in Fig.\ref{fig9a}--\ref{fig9f} for both skin and fat tissue layers. As shown in Fig.\ref{fig9c} and \ref{fig9d}, the X-directed slot reduces the SAR value compared to the reference antennas of Fig.\ref{fig9a} and \ref{fig9b}. The nature of distribution is also changed as there appears a concentric maxima around the centre for the X-directed slot. However, for the Y-directed slot, other constitutive parameters like $\sigma$ and $\rho$ also control the variation of SAR following \eqref{eq12}. As a consequence, the SAR value is enhanced  by the Y-slot for the skin layer where as for the fat layer, it is slightly increased. Comparison of the maximum SAR values of the line fed antennas have been reported in Table \ref{Table3}. The impacts of X-directed slot is found to be maximum for the microstrip line feeding with respect to the probe feeding because the direction of the current in the feed line is parallel to the target Y-polarised edge mode of the antenna. So the modal weighting coefficient $\alpha_n$ of \eqref{5} starts to be an influencing parameter. As two vectors $[V]$ and $[I_m]$ becomes parallel, their inner product holds higher value in the line fed antennas. Thus, both the radiating element's modal characteristics and the orientation of the excitation signal control the resultant radiation behaviour.  

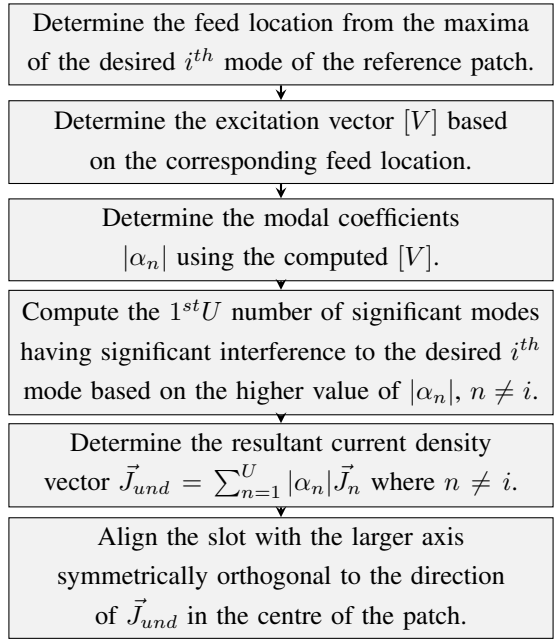
\begin{figure}[ht!]
\centering
\begin{tikzpicture}[node distance=1.5cm][font=\small]

\node (pro1) [process] {Determine the feed location from the maxima of the desired $i^{th}$ mode of the reference patch.};

\node (pro2) [process, below of=pro1, yshift=0.2cm] {Determine the excitation vector $[V]$ based on the corresponding feed location.};

\node (pro3) [process, below of=pro2, yshift=0.2cm] {Determine the modal coefficients $|\alpha_n|$ using the computed $[V]$.};

\node (pro4) [process, below of=pro3] {Compute the $1^{st} U$ number of significant modes having significant interference to the desired $i^{th}$ mode based on the higher value of $|\alpha_n|$, $n \neq i$.};

\node (pro5) [process, below of=pro4] {Determine the resultant current density vector $\vec{J}_{und}= \sum_{n=1}^U
|\alpha_n|\vec{J}_n$ where $n \neq i$.};

\node (pro6) [process, below of=pro5] {Align the slot with the larger axis symmetrically orthogonal to the direction of $\vec{J}_{und}$ in the centre of the patch.};

\draw [arrow] (pro1) -- (pro2);
\draw [arrow] (pro2) -- (pro3);
\draw [arrow] (pro3) -- (pro4);
\draw [arrow] (pro4) -- (pro5);
\draw [arrow] (pro5) -- (pro6);
\end{tikzpicture}
 \caption{Procedure for determining the orientation of the narrow slot.}
 \label{fig:flow1}
\end{figure}
 
In general, the characteristic mode analysis is found to provide systematic guideline for maintaining the radiated signal's polarization requirement and SAR distribution along a desired direction. For arbitrary orientation of the current distribution, the procedure of slot loading has been included in Fig. \ref{fig:flow1}.
\newpage
\section{Conclusion}
This work explored the impacts of slot loading in rectangular patch antennas. Initially characteristic mode analysis of the radiating structures were performed to understand the perturbation in the current and field distribution. Later full-wave study had been carried out to investigate the loading effects on the electromagnetic energy absorption by biological tissues placed in the near field region. It was found that the prior knowledge of the characteristic modes of the radiating geometry can provide suitable information about the arrangement of the narrow slot for reducing the SAR value of the neighbouring tissues. The orientation of the slot is found to determine the amount of maximum radiation in a particular direction. Such important feature of the CMA can be further utilised for focussing the  power of the wireless power transfer system to a target region of interest.\\


\bibliographystyle{IEEEtran}
\bibliography{IEEEabrv,Ref_JERM}

\end{document}